\documentclass{ifacconf}

\usepackage{graphicx}      % include this line if your document contains figures
\usepackage{natbib}        % required for bibliography

\usepackage{amsmath,amssymb,amsfonts}

\usepackage{xcolor}
\usepackage{xspace}
\usepackage{soul}
\usepackage{mathtools}
\usepackage{makecell}
\usepackage{nicefrac}
\usepackage{tcolorbox}

\newcommand{\ra}{\rightarrow}

\newcommand{\ie}{\unskip, i.\,e.,\xspace}
\newcommand{\eg}{\unskip, e.\,g.,\xspace}

\newcommand{\R}{\ensuremath{\mathbb R }}

\newcommand{\sm}{\ensuremath{\setminus}}

\newcommand{\nat}{\text{na}}
\newcommand{\car}{\text{oc}}
\newcommand{\nit}{\text{ni}}
\newcommand{\m}{\text{mi}}

\newcommand*\diff{\mathop{}\!\mathrm d }

\definecolor{dgreen}{rgb}{0.0, 0.5, 0.0}

\newcommand{\norm}[1]{\left\lVert#1\right\rVert}  % vector norm, needs ams
\newcommand{\abs}[1]{\left\lvert#1\right\rvert}

\newcommand{\set}[1]{\mathbb #1}

\usepackage{url}
\usepackage{tikz}
\usetikzlibrary{arrows,%
                plotmarks, calc}
\usetikzlibrary[positioning, calc, intersections, backgrounds]

\begin{document}
\begin{frontmatter}

\title{High-gain observer for the nitrification process including sensor dynamics} 

\thanks[footnoteinfo]{This work was partially funded by the European Union, European Social Found ESF, Saxony. \\
©\the\year \ the authors. This work has been accepted to IFAC for publication under a Creative Commons Licence CC-BY-NC-ND.}

\author[TUC]{Patrick Schmidt, Arne-Jens Hempel, Stefan Streif} 

\address[TUC]{Technische Universit\"at Chemnitz, 09126 Chemnitz, Germany, Automatic Control and System Dynamics Lab (e-mail: \{patrick.schmidt, arne-jens.hempel, stefan.streif\}@etit.tu-chemnitz.de)}

\begin{abstract}                % Abstract of not more than 250 words.
Fertilization is commonly used to increase harvests.
The lack of knowledge of soil properties and the excessive use of fertilizers can result in overfertilization.
Current sensor technology is able to measure the concentrations of some of the involved substances only at selected locations and depths.
Point measurements of adjacent sensors in coarse sensor networks can be used to infer upon the state of nitrate concentrations in the sensor surroundings. 	
For this purpose, a high-gain observer is proposed. 
Models of the nitrification process as well as the measurement dynamics for the observer design are derived and discretized on a grid to obtain a system of ordinary differential equations. 
It is shown that the nonlinearities of the model can be bounded and how the observer gain can be computed via linear matrix inequalities.
Furthermore, a model reduction is proposed, which allows the consideration of more grid points.
A simulation study demonstrates the proposed approach.% as prove of concept.
\end{abstract}


\end{frontmatter}
%===============================================================================

\section{Introduction} \label{sec:intro}

%Nowadays, food supply for an increasing global population is a huge and an important task.
Fertilization is used to increase harvest and ensures a stable supply with fruits, vegetables and cereals.
%Most commonly, the amount of fertilizers is used heuristically, which often leads to overfertilization.
%In addition uncertainties about the already existing nutrients in soil may also cause an overfertilization} \citep{Innes2013-economics}.
Because the amounts of nutrients in the soil are not known precisely, the amount of fertilizers to be applied is determined heuristically, which often leads to overfertilization \citep{Innes2013-economics}.

Overfertilization comes along with lots of negative long- and short-term consequences and is inflicting different ecosystems, as clarified in \citep{Albornoz2016-crop, Basso2016-environmental, Cui2010-season, Fernandez2000-effect, Weinbaum1992-causes}. 
%\orange{For example it adversely affects flowering, fruit set, and fruit growth \citep{Weinbaum1992-causes}.
%As also investigated, overfertilization causes pathological and physiological disorders as well as susceptibility to disease and insect pests in orchard crops \citep{Weinbaum1992-causes}. 
%Especially, nitrogen overfertilization causes adulterated flavor, reduced content of other mineral nutrients and a reduced sythesis of vitamins \citep{Albornoz2016-crop}. 
%Further negative impacts were also studied on olive trees, where polyphenol content, bitterness and decreasing oil stability was investigated \citep{Fernandez2000-effect}.}

In addition to direct effects, the following indirect effects were also recorded: pollution of air, soil, and especially ground water \citep{Savci2012-investigation}. 
Only half of the fertilized amount of nitrogen, nitrate and other components is absorbed by plants.
The surplus sinks through soil and enters ground water decreasing the quality of water in general.
The consequences are far-reaching; extinction of animal and plant species, human diseases in less-developed countries, rising costs for processing and providing drinking water, just to mention a few \citep{Savci2012-investigation}.
%Beside nitrate, also cadmium, phosphorius and even uranium are contained in fertilizers \citep{Savci2012-investigation} and arrive in ground water.

Due to these negative consequences of overfertilization, the European Union aims to control fertilization and improve water quality, which was first presented in the \emph{Nitrate Directive} in 1991 \citep{EU2010-nitrate}.
Since then, the total amount of mineral fertilizers was reduced; however, the total consumption of nitrogen fertilizer has increased \citep{EU2022-nitrate}.
%For example in 2018 the amount of nitrogen fertilizer in the EU was about 10.2 million tonnes, which is an increased of 1.9 percent in comparison to 2008, whereas phosphorus fertilizer was reduced to 1.1 million tonnes, or 1.2 percent less than in 2008 \citep{EU2020-nitrate}.
%Therefore, the reduction of nitrogen in fertilization remains an open issue.
In this context, farming remains responsible for the most part of total nitrogen discharge into surface water. 
Accordingly, on-site measurements of different fertilizer components are an important step to control fertilization.
%, which is is not only beneficial for the environment and the associated climate change, but also for monetary reasons.
%Hence it remains an important challenge to decrease the use of nitrogen, especially in agriculture \citep{EU2010-nitrate}.
%Especially in Germany, the amount of nitrate in ground water is higher than in most of the other EU member states.
%For this reason, the EU imposed penalties to improve the quality of ground water \citep{EU2018-nitrate}.
%Obviously, the efficient and responsible use of fertilizers is an important and relevant aim.

%Since all the named factors affect human life, a surveillance based on measurements of different fertilizer components could be a first step to control fertilization, which is is not only beneficial for the environment and the associated climate change, but also for monetary reasons.
There exist various approaches for measuring the concentration of nitrate and other substances, both in soil and in water, e.g. standard chemical analysis \citep{Ali2017-microfluidic, Moo2016-new, Wang2017-methods}. 
However, for this paper we consider the application of wireless electrochemical sensors for in situ measurements.
Those provide a ``real-time'' measurement and can be placed freely, but have the following limitations. 
%A very common example are electrochemical sensors, which consist of at least two electrodes and an electrolyte in between.
%If the sensor detects the substance, an electrochemical reduction occurs at one electrode.
%This reaction generates a voltage difference that is measured.
%Based on this electrochemical reaction on the sensor, the detection and identification of nitrate ions is possible \citep{Bakker2002-electrochemical}.
%As a disadvantage, materials on the surface wear during the measurement, which may create measurement errors.
First, not every kind of substances can be measured by sensors due to technical limitations.
Second, concentrations can only be measured on-spot at the location of the sensor.

To identify the distribution of all relevant concentrations despite these difficulties, a set of sensors is arranged on the considered field. 
These sensors are then connected and synchronized to a network in order to transmit the measurements and form the basis for estimation.

\begin{figure}[!h]
	\begin{center}
	\includegraphics[width =0.45\textwidth]{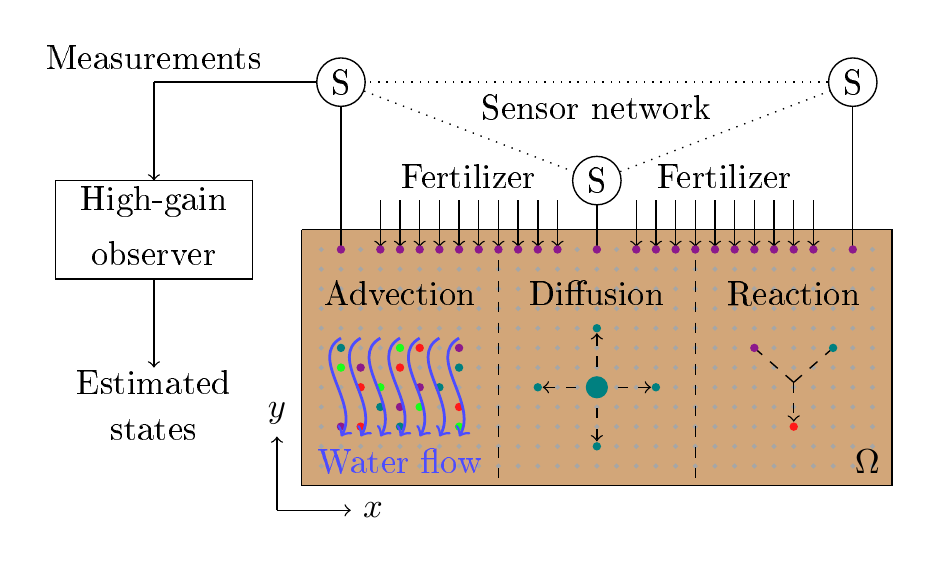}
	\caption{Visualization of the setup: Nitrate (violet) gets into soil via fertilization and reacts with organic carbon (teal) under the presence of microbes (green) to nitrite (red). All substances are transported through soil via advection and diffusion. 
	A network of sensor nodes measures nitrate concentrations. Based on these measurements, a high-gain observer estimates all the other concentrations.}
	\label{fig:setup}
		\end{center}
\end{figure}

In the scope of the paper, this estimation is realized by an observer which itself requires a model.
This model is the nitrification process, which describes the reaction of nitrate and organic carbon to nitrite under the influence of microbes operating as a catalyst \citep{Bernhard2010-nitrogen, Stein2016-nitrogen}.
The transport of these substances in soil is governed by the advective transport by the flow field and the molecular diffusion \citep{Cai2013-numerical}, yielding a system of partial differential equations (PDEs). 
The introduced setting is summarized and depicted in Fig.~\ref{fig:setup}.

There exist results on observer design for PDEs %\citep{Cornilleau2012-controllability, Elamvazhuthi2018-bilinear, Georges1995-use, Vaidya2012-actuator}
\citep{Ahmed2015-adaptive, Kharkovskaia2017-design, Schaum2014-matrix}, but their assumptions and requirements cannot be fulfilled within the scope of our research.  
Furthermore, their application is more complicated in comparison to observers for ordinary differential equations (ODEs).
Therefore, a discretization of the PDE to a system of ODEs is considered, which delivers the models for the observer design.

There are many design approaches for observers of nonlinear systems \citep{Besanccon2007-nonlinear, Schaffner1999-variants, Wan2000-unscented, Zeitz1987-extended}. 
Some of them require a special structure of the system dynamics such as the observer canonical form \citep{Robenack2007-high}.
The transformation of the given system into a canonical form might be challenging, since it can be accompanied by complex calculation such as Lie-derivatives.
In contrast to the mentioned works, high-gain observer design is characterized by a simple approach \citep{Khalil2008-high, Robenack2012-structure}.

Within this contribution, we show how such a high-gain observer can be applied to infer on concentrations in the sensors vicinity. 
Those high-gain observers are based on solving a (high-dimensional) linear matrix inequality.
Depending on the structure of the matrices, this task may be accompanied by a high computational effort.
Therefore, a reduction of the underlying model is a suitable way to reduce this effort, which is also discussed in the following chapters.
Once the concentrations are computed, they can be used to determine the necessary amount of fertilizer.
For example, \citep{Russwurm2020-optimal} presents three different approaches to obtain a desired fertilizer distribution on a field, which are based on the known amount of fertilizer at different points of a grid.

The paper is organized as follows.
The problem is described in Section \ref{sec:model}, where the underlying model of the nitrification process is deduced and discretized on a two-dimensional grid to obtain a system of ODEs.
Afterwards, a high-gain observer is designed for the derived model in Section \ref{sec:methods}, which is reformulated to satisfy the necessary structure.
Thereby, observability is investigated and an additional coupling between the different substances is introduced to ensure it.
Furthermore, the nonlinearity of the system is bounded to apply the high-gain observer.
As a last point in this section, a reduction of the model is proposed to decrease the dimension of the system.
Section \ref{sec:impl-case-study} considers the implementation of two different high-gain observers and their application at the system.
Furthermore, a model reduction is presented.
Finally, Section \ref{sec:concl-outlook} summarizes the results of the paper and gives an outlook on some possible extensions.

\section{Modeling and Problem Setting} \label{sec:model}

The application of a functional observer requires a dynamic model formalizing the behavior and distribution of the four substances and sensors involved\ie for nitrate (na), organic carbon (oc), nitrite (ni), and microbes (mi).
A capable model type of the nitrification process is the advection-diffusion-reaction model \citep{Cai2013-numerical}
\begin{equation} \label{eqn:ADR}
	\frac{\mathrm d}{\mathrm d t} c = \underbrace{\Gamma \nabla^2 c}_{\text{diffusion}} - \underbrace{v^\top \nabla c}_{\text{advection}} + \underbrace{r(c^{\text{acc}}, c^{\text{don}}, c^{\m})}_{\text{reaction}}
\end{equation}
on a set $\Omega \subset \R^2$ with initial condition $c(x,0) = c_0(x) \in \R, x \in \Omega$ and boundary condition $c(x,t) = \tilde c(x) \in \R \text{ on } \partial \Omega \times [0, T]$.
It is now considered in detail and tailored to the task at hand.
The right-hand side of \eqref{eqn:ADR} contains three terms, i.e. diffusion, advection, and, reaction, as shown in Fig. \ref{fig:setup}.

For the considered application, the diffusion term describes the movement of a substance with concentration $c \in \R$ from regions with a higher concentration to regions with a lower concentration, where $\Gamma > 0$ is given as molecular diffusivity of the corresponding concentration.
Diffusion occurs for all four involved concentrations \ie nitrate, organic carbon, nitrite, and microbes and it describes their movement on a slice of the field notated as $\Omega \subset \R^2$.

The advection term describes the transport of a substance with concentration $c \in \R$ via a velocity field $v$.  
Since a slice of field $\Omega \subset \R^2$ is considered, $v$ is given as $v = \begin{pmatrix} v^{\text{x}} & v^{\text{y}} \end{pmatrix}^\top \in \R^2$, with $v^{\text{x}}$ as the velocity in $x$ direction and $v^{\text{y}}$ as the velocity in $y$ direction.
Convection occurs also for all four concentrations, since all of them are transported by water.

The reaction term depends on the concentrations of the electron acceptor $c^{\text{acc}}$, the electron donor $c^{\text{don}}$, and also on the concentration of microbes $c^{\m}$. 
It may have negative sign depending on whether educts or products are considered.
%The reaction term decreases the concentration of educts, but it increases the concentration of the products, which is visualized by $\pm$ in \eqref{eqn:ADR}.
In general, this reaction can be described by the Monod-equation \citep{Cheng2016-reactive}
\begin{equation} \label{eqn:Monod}
	r(c^{\text{acc}}, c^{\text{don}}, c^{\m}) = r_{\text{max}} \frac{c^{\text{acc}}}{K_{\text{acc}} + c^{\text{acc}}} \frac{c^{\text{don}}}{K_{\text{don}} + c^{\text{don}}} c^\m,
\end{equation}
with a maximum growth rate of micro-organisms $r_{\text{max}} > 0$ and constants $K_{\text{acc}} > 0$ and $K_{\text{don} > 0}$ as half-velocity-constants of the electron acceptor and the electron donor.
In the nitrification process, the electron acceptor is nitrate and the electron donor is organic carbon.
Therefore, the constants in \eqref{eqn:Monod} are denoted as $K_{\text{acc}} = K_\nat$ and $K_{\text{don}} = K_\car$.
During reaction both concentrations as well as microbes decrease, whereas nitrite as the product of the reaction increases.

As a result, the nitrification process can be modeled as
\begin{equation} \label{eqn:PDE-system}
	\begin{split}
		\frac{\mathrm d}{\mathrm d t} c^\nit &= \Gamma \nabla^2 c^\nit - v^\top \nabla c^\nit + r(c^\nat, c^\car, c^\m), \\
		\frac{\mathrm d}{\mathrm d t} c^i &= \Gamma \nabla^2 c^i - v^\top \nabla c^i - r(c^\nat, c^\car, c^\m), \\
	\end{split}
\end{equation}
for $i \in \{ \nat, \car, \m \}$ with concentrations $c^\nat, c^\car, c^\nit, c^\m \in \R$ of the four substances.
The molecular diffusivity $\Gamma$ is assumed to be the same for each substance.

After modeling the transport and the reaction of the involved substances, the dynamics of the sensors will be included.
As it is visualized in Fig. \ref{fig:setup}, the sensors are distributed on the field $\Omega$ and are able to measure the concentrations of nitrate at their specific location.
The dynamics of one sensor is assumed as a 1$^\text{st}$-order model:
\begin{equation} \label{eqn:PT1-nat-nit}
	\frac{\mathrm d}{\mathrm d t} \bar c^\nat = \frac{1}{\tau_\nat} (c^\nat - \bar c^\nat),
\end{equation}
with time constant $\tau_\nat$ and $\bar c^\nat$ as the measurement of $c^\nat$.

The combination of \eqref{eqn:PDE-system} and \eqref{eqn:PT1-nat-nit} yields the model of the measured nitrification process with the involved substances of nitrate, organic carbon, microbes and nitrite.
Since nitrite is only the product of this chemical reaction (cf. \eqref{eqn:PDE-system}), it does not matter for the amount of nitrate in soil.
Therefore, the nitrite concentrations are excluded to reduce the number of equations.
The model is given as
\begin{equation} \label{eqn:system-cont-time}
	\begin{split}
		\frac{\diff}{\diff t} \begin{pmatrix} c^\nat \\ c^\car \\ c^\m \\ \bar c^\nat  \end{pmatrix} &= \begin{pmatrix}
			\Gamma \nabla^2 c^\nat - v^\top \nabla c^\nat - r(c^\nat, c^\car, c^\m) \\
			\Gamma \nabla^2 c^\car - v^\top \nabla c^\car - r(c^\nat, c^\car, c^\m) \\
			\Gamma \nabla^2 c^\m - v^\top \nabla c^\m - r(c^\nat, c^\car, c^\m) \\
			\frac{1}{\tau_\nat} (c^\nat - \bar c^\nat)
		\end{pmatrix}, \\
		\bar c^\nat &= \begin{pmatrix}
			0 & 0 & 0 & 1
		\end{pmatrix} \begin{pmatrix} c^\nat \\ c^\car \\ c^\m \\ \bar c^\nat \end{pmatrix},
	\end{split}
\end{equation}
with
\begin{itemize}
	\item states: $c^\nat, c^\car, c^\m$, $\bar c^\nat$
	\item measurable states: $\bar c^\nat$, and
	\item parameters: $K_\nat$, $K_\car$, $r_{\text{max}}$, $\Gamma$, $\tau_\nat$, $v^\text{x}$, $v^{\text{y}}$.
\end{itemize}
To reduce overfertilization and to apply the required amount of fertilizer, it is necessary to determine the non-measurable concentrations on the field.
Which is conceived as an estimation task, for which we want to apply a high-gain observer, requiring a system of ODEs.
Therefore, \eqref{eqn:system-cont-time} is discretized on a two-dimensional grid.

\subsection{Discretization of the PDE}

The PDE \eqref{eqn:system-cont-time} is discretized equidistantly in space,
%The discretization of \eqref{eqn:system-cont-time} is conducted equidistantly in space
 which yields a grid of $N + 2$ points in each direction. 
The distance between two adjacent grid points is given by the edge length $\Delta$.
As a short hand notation $c_{j,k}$ is used for a concentration in column $j$ and row $k$ of the grid, $j, k \in \{0, 1, \ldots, N +1 \}$.
The points $\{ c_{j,k}: j, k \in \{ 0, 1, \ldots, N + 1 \} \} \sm \{ c_{j,k}: j, k \in \{ 1, 2, \ldots, N \} \}$ lie on the boundary $\partial \Omega$ of $\Omega$.
This notation holds for all concentrations, as well as velocities $v^{\text{x}}_{j,k}$ and $v^{\text{y}}_{j,k}$.

Gradients and Laplace-operators in \eqref{eqn:system-cont-time} are replaced at the points $c_{j,k}, j, k \in \{1, 2, \ldots, N \}$ in the interior of $\Omega$ by finite-difference method according to \citep{Nagel2011-solving}
\begin{equation} \label{eqn:replacement-laplace-gradient}
	\begin{split}
		v^\top \nabla c &\mapsto v^{\text{x}}_{j,k} \frac{c_{j,k} - c_{j - 1,k}}{\Delta} + v^{\text{y}}_{j,k} \frac{c_{j,k} - c_{j,k - 1}}{\Delta}, \\
		\nabla^2 c &\mapsto \frac{c_{j + 1,k} + c_{j,k + 1} + c_{j - 1,k} + c_{j,k - 1} - 4 c_{j,k}}{\Delta^2}.
	\end{split}
\end{equation}
The vertices on the border of the discretization grid $\{ c_{j,k}: j, k \in \{ 0, N + 1 \} \}$ do not appear in \eqref{eqn:replacement-laplace-gradient}.
They might be treated as inputs, since their value is increased either by fertilization on the surface or by the concentrations of an adjacent field.
In our case, periodic boundary conditions are considered\ie at the left-hand side and the right-hand side of the field we assume $c^i_{0,k} = c^i_{N,k}$ and $c^i_{N + 1,k} = c^i_{1,k}$ for $i \in \{ \nat, \car, \m \}$ and $k = 1, 2, \ldots, N$.
At the lower boundary, $c_{j,N + 1} = c_{j,N}$, $j = 1, 2, \ldots, N$ is assumed to hold \ie the points the this boundary have the same concentration as the the points in the interior above them.
Finally, the remaining $N$ points at the upper boundary serve as inputs and they are stacked into
\begin{equation}
	C^i_\partial = \left(c_{1,0}^i \: c_{2,0}^i \cdots c_{N,0}^i \right)^\top \in \R, \ i \in \set I := \{ \nat, \car, \m \}.
\end{equation}
Their values are increased by fertilization on the surface.

The $N^2$ concentrations in the interior of $\Omega$ are arranged in a similar way:
\begin{equation}
	C^i = \left(c_{1,1}^i \: c_{1,2}^i \cdots c_{1,N}^i \: c_{2,1}^i \cdots c_{N,N}^i \right)^\top \in \R^{N^2}, \ i \in \set I,
\end{equation}
which means that $c_{j,k}^i$ is the $(j-1)N + k$-th element of $C^i$.
Let $p$ be the number of sensors measuring nitrate.
At each of the $p$ positions of the network where a sensor is placed, the dynamics of the sensor is given as \eqref{eqn:PT1-nat-nit}.
The $p$ measured values of nitrate are stacked in $\bar C^\nat \in \R^p$.
They are formalized with the help of a $p \times N^2$ logical selection matrix $M$, encoding the $p$ sensor positions for the corresponding vectors of concentration $C^\nat$ as $\bar C^\nat = M C^\nat$.
The concentrations of nitrate, organic carbon, and microbes, as well as the measurements are combined to the state vector $x \in \R^n, n = 3 N^2 + p$, the input vector $u \in \R^m, m = 3N$ and the output vector $y \in \R^{p}$ ($p$ defined later) according to
\begin{equation} \label{eqn:choice-of-x-u-y}
	x = \begin{pmatrix} C^\nat \\ C^\car \\ C^\m \\ \bar C^\nat \end{pmatrix}, 
	u = \begin{pmatrix} C^\nat_\partial \\ C^\car_\partial \\ C^\m_\partial \end{pmatrix}, 
	y = \bar C^\nat.
\end{equation}

The discretization leads to the following equations for each point in the interior of $\Omega$:
\begin{equation} \label{eqn:system-discr-space}
	\begin{split}
		\dot x &= %\frac{\diff}{\diff t} \begin{pmatrix} C^\nat \\ C^\car \\ \bar C^\nat \\ C^\nit \\ C^\m \\ \bar C^\nit \end{pmatrix} = 
			\begin{pmatrix}
			f_{\mathrm c}(C^\nat, C^\nat_\partial) - R(C^\nat, C^\car, C^\m) \\
			f_{\mathrm c}(C^\car, C^\car_\partial) - R(C^\nat, C^\car, C^\m) \\
			f_{\mathrm c}(C^\m, C^\m_\partial) - R(C^\nat, C^\car, C^\m) \\
			\dfrac{1}{\tau_\nat} (M C^\nat - \bar C^\nat)
			\end{pmatrix}, \\
		y &= %\begin{pmatrix} \bar C^\nat \\ \bar C^\nit \end{pmatrix} = 
			\underbrace{\begin{pmatrix}
			0_{p \times 3 N^2} & I_{p}	\end{pmatrix}}_{= C} x
	\end{split}
\end{equation}
where $I_p$ is the $p \times p$ identity matrix and $0_{p \times N^2}$ is the $p \times N^2$ zero matrix.
To avoid confusion, $C$ is given as the output matrix, while all concentrations are written with a superscript $i \in \set I$.
Furthermore, $f_{\mathrm c}: \R^{N^2} \times \R^{N} \ra \R^{N^2}$, as well as $R: \R^{N^2} \times \R^{N^2} \times \R^{N^2} \ra \R^{N^2}$ are vector-valued mappings.
They are defined element-wise as
\begin{equation} \label{eqn:system-discr-space-fc-reac}
	\begin{split}
		&f_{\mathrm c}(c^i_{j,k}, C^i_\partial) \\
		&= \Gamma \frac{c^i_{j + 1,k} + c^i_{j,k + 1} + c^i_{j - 1,k} + c^i_{j,k - 1} - 4 c^i_{j,k}}{\Delta^2} \\
		& \ \ \ - v^{\text{x}}_{j,k} \frac{c^i_{j,k} - c^i_{j - 1,k}}{\Delta} - v^{\text{y}}_{j,k} \frac{c^i_{j,k} - c^i_{j,k - 1}}{\Delta}, \\
		&R(c^\nat_{j,k},c^\car_{j,k},c^\m_{j,k}) = r_{\text{max}} \frac{c^\nat_{j,k}}{K_\nat + c^\nat_{j,k}} \frac{c^\car_{j,k}}{K_\car + c^\car_{j,k}} c^\m_{j,k}, \\
		&i \in \set I , \ j,k \in \{ 1, 2, \ldots, N \}.
	\end{split}
\end{equation}

\section{Methods} \label{sec:methods}

In this section, the derived model is used for the design of an observer.
Since only some concentrations are measured on the field due to the limitations that were described in the introduction, the observer is used to estimate the state vector $x$ based on inputs $u$ and measurements $y$ given in \eqref{eqn:choice-of-x-u-y}.
In the following, the high-gain observer is introduced to complete the task.

\subsection{high-gain observer} \label{sec:highgainobserver}
%Systemstruktur
Generally, high-gain observers require a system structure of the following type: 
\begin{equation} \label{eqn:highgain-system}
	\begin{split}
	\dot x &= A x + \tilde \Phi(x, u) \\ 
	y &= C x
	\end{split}
\end{equation}
with matrices $A \in \R^{\tilde n \times \tilde n}$ and $C \in \R^{\tilde p \times \tilde n}$, as well as a nonlinear map $\tilde \Phi: \R^{\tilde  n} \times \R^{\tilde m} \ra \R^{\tilde n}$.
According to \citep{Thau1973-observing} the observer structure for \eqref{eqn:highgain-system} is set to
\begin{equation} \label{eqn:highgain-observer}
	\begin{split}
	\dot{\hat x} &= A \hat x + \tilde \Phi(\hat x, u) + L (y - \hat y), \\
	 \hat y &= C \hat x.
	\end{split}
\end{equation}
Applying the observer \eqref{eqn:highgain-observer} on the system \eqref{eqn:highgain-system} yields the observation error dynamics for $e := x - \hat x$ as
\begin{equation} \label{eqn:highgain-error}
	\dot{e} = (A - L C) e + \tilde \Phi(x, u) - \tilde \Phi(\hat x, u).
\end{equation}
For the application at hand, the system structure given in \eqref{eqn:system-discr-space} needs to meet the necessary structure \eqref{eqn:highgain-system}.
Inserting \eqref{eqn:system-discr-space-fc-reac} into \eqref{eqn:system-discr-space} yields the following structure:
\begin{equation} \label{eqn:system-ZRD}
	\begin{split}
		\dot{\hat x} &= A \hat x + B u + \Psi(\hat x),\\
		\hat y &= C \hat x, \\
		\Psi(\hat x) &= \begin{pmatrix}
			-R(C^\nat, C^\car, C^\m) \\ -R(C^\nat, C^\car, C^\m) \\ -R(C^\nat, C^\car, C^\m) \\ 0_{p \times 1}
		\end{pmatrix},
	\end{split}
\end{equation}
where $\hat x, u$ are given in \eqref{eqn:choice-of-x-u-y} and matrices $A \in \R^{n \times n}$ and $B \in \R^{n \times m}$,  $n =  3 N^2 + p, m = 3 N$ are obtained by the measurement dynamics \eqref{eqn:PT1-nat-nit} and the advection-diffusion term $f_{\mathrm c}(\cdot)$ according to \eqref{eqn:system-discr-space-fc-reac}.

Besides the necessary system structure, there are further conditions that need to be met to ensure the convergence of the error dynamics.
Namely, these are the observability and the boundedness condition, which are analyzed subsequently.

\subsection{Observability Condition} \label{sec:observ}
To ensure asymptotic stability of \eqref{eqn:highgain-error}, the observer gain $L$ is designed such that the nonlinearities are dominated by the linear part \citep{Robenack2012-structure}.
Since $L$ has to be chosen in order to make the dynamics of $\dot{e} = (A - L C) e$ arbitrarily fast, the the pair $(A, C)$ in \eqref{eqn:system-ZRD} has to be observable.
Furthermore, the nonlinear part has to be bounded, since this ensures the dominance of the linear part.
At first, observability is checked via the Hautus criterion. 

Since the advection and diffusion terms only model the transport of the respective concentrations, the matrix $A$ is given in block-diagonal structure.
Due to this structure of $A$, the pair $(A,C)$ is remains unobservable for any task-specific $C$ \ie only some concentrations can be measured by sensors.
 
Consequently, there is no coupling between different concentrations in the linear part, in particular microbes and organic carbon.
Therefore, we propose the following slight modification of \eqref{eqn:system-ZRD} 
\begin{equation} \label{eqn:coupling}
	\begin{split}
		\dot{\hat x} &= A \hat x + B u + \Psi(\hat x) \\
		&= [A - \tilde A] \hat x + \underbrace {[\tilde A \hat x + B u + \Psi(\hat x)]}_{= \Phi(\hat x,u)} ,
	\end{split}
\end{equation}
where $\tilde A$ is chosen such that $(A - \tilde A, C)$ is observable.
The nonlinearity of \eqref{eqn:coupling} is defined as $\Phi: \R^n \times \R^m \ra \R^n$, which yields the following observation error dynamics:
\begin{equation} \label{eqn:system-highgain-error}
	\dot{e} = (A - \tilde A - L C) e + \Phi(x, u) - \Phi(\hat x, u).
\end{equation}
For the application at hand $\tilde A$ is set to include couplings between $C^\nat$, $C^\car$, and $C^\m$.
As a natural linkage, we consider the linearization of $\Psi(x)$, such that
\begin{equation} \label{eqn:artifical-coupling}
	\tilde A := \nabla_x \Psi(x)\mid_{x = x_{\text{OP}}},
\end{equation}
where $x_{\text{OP}}$ is the operating point of the observer.
Since coupling different concentrations is more important than a concrete value, the choice of the operating point is not crucial for the following considerations.

The number of sensors $p$ is chosen such that $C^\nat$ can be estimated based on the measurement of $\bar C^\nat$ and the inputs.
It turned out that three sensors are sufficient to satisfy observability of a system including only one concentration. 
A detailed analysis of this result is beyond the scope of this paper.

Since observability of the linear part is ensured due to the additional coupling, an observer gain $L$ can be computed.
Based on the chosen approach of the high-gain observer, a Lipschitz constant as a bound for the nonlinearity appears in its computation.
Therefore, this bound of the nonlinear term is considered in the next subsection.

\subsection{Bounding the Nonlinear Reaction Term} \label{sec:nonlinear}
To ensure a dominant linear dynamics in \eqref{eqn:system-highgain-error}, the nonlinearity has to be bounded.
This bound can be estimated with the help of a Lipschitz constant which is computed for \eqref{eqn:coupling} in the following.

A Lipschitz constant $\gamma \in \R_+$ of the nonlinearity of \eqref{eqn:coupling} has to satisfy
\begin{equation}\label{eqn:Lipschitz_nonlin}
	\norm{\Phi(w, u) - \Phi(z, u)} \leq \gamma \norm{w - z}
\end{equation} 
for all $w,z \in \R^n$.
Since 
\begin{equation}
	\begin{split}
		&\norm{\Phi(w, u) - \Phi(z, u)} \\
		&= \norm{\Psi(w) + \tilde A w - \Psi(z) - \tilde A z} \\
%		&\leq \norm{\Psi(w) - \Psi(z)} + \norm{ \tilde A (w - z)} \\
		&\leq \norm{\Psi(w) - \Psi(z)} + \norm{\tilde A}_{\text{F}} \norm{w - z},
	\end{split}
\end{equation}
the Lipschitz constant for the nonlinear map $\Psi(\cdot) = \begin{pmatrix} \psi_1(\cdot) & \psi_2(\cdot) & \cdots & \psi_n(\cdot) \end{pmatrix}^\top$ remains to be identified.
Therefore, let $w = \begin{pmatrix}
	w_1^\top & w_2^\top & \cdots & w_{N^2}^\top
\end{pmatrix}^\top$ be given, where $w_j = \begin{pmatrix}
	c_j^\nat & c_j^\car & c_j^\m
\end{pmatrix}^\top \in \R^3$ and $c_j^i$ is the $j$-th element of $C^i$, $i \in \{ \nat, \car, \m \}$.
Vector $z$ is defined analogously. 
The considered term can be bounded by 
\begin{subequations}
	\begin{align}	
		&\norm{\Psi(w) - \Psi(z)} %= \sqrt{\sum_{i = 1}^n (\psi_i(w) - \psi_i(z))^2} \\
		=\sqrt{3 \sum_{i = 1}^{N^2} (\psi_i(w) - \psi_i(z))^2} \label{eqn:Lipschitz_nonlin_proof1} \\
		&\leq \sqrt 3 \sum_{i = 1}^{N^2} \sqrt{(\psi_i(w) - \psi_i(z))^2} = \sqrt 3 \sum_{i = 1}^{N^2} \underbrace{\abs{\psi_i(w) - \psi_i(z)}}_{\leq \varrho \norm{w_i - z_i}} \label{eqn:Lipschitz_nonlin_proof2} \\
		&\leq \sqrt 3 \varrho \sum_{i = 1}^{N^2} \norm{w_i - z_i} = \sqrt 3 \varrho \sqrt{ \left( \sum_{i = 1}^{N^2} \norm{w_i - z_i}\right)^2} \label{eqn:Lipschitz_nonlin_proof3} \\
		&\leq \sqrt 3 \varrho \sqrt{ N^2 \sum_{i = 1}^{N^2} \norm{w_i - z_i}^2} = \sqrt 3 N \varrho \norm{w - z}. \label{eqn:Lipschitz_nonlin_proof4}
	\end{align}
\end{subequations}
Equality \eqref{eqn:Lipschitz_nonlin_proof1} is obtained, since the reaction terms in \eqref{eqn:system-ZRD} are the same.
Inequality \eqref{eqn:Lipschitz_nonlin_proof2} is obtained since the square root is monotone, whereas \eqref{eqn:Lipschitz_nonlin_proof3} is obtained via Lipschitz continuity of the functions $\psi_i(\cdot)$, which are given as the reaction terms of the respective concentrations.  
Since $\psi_i$ is independent of all $w_j, z_j, j \not= i$, the term $\abs{\psi_i(w) - \psi_i(z)}$ can be bounded via $\varrho \norm{w_i-z_i}$, since all other values in $w$ and $z$ would yield a higher Lipschitz-constant. 
In our case, $\varrho$ was computed numerically as
\begin{equation}
	\varrho := \text{max}_{(c^\nat_{\cdot,\cdot}, c^\car_{\cdot,\cdot}, c^\m_{\cdot,\cdot})^\top \in \R^3} \norm{\nabla_{(c^\nat_{\cdot,\cdot}, c^\car_{\cdot,\cdot}, c^\m_{\cdot,\cdot})^\top} R(c^\nat_{\cdot,\cdot}, c^\car_{\cdot,\cdot}, c^\m_{\cdot,\cdot})}.
\end{equation}
Finally, \eqref{eqn:Lipschitz_nonlin_proof4} is obtained via a generalization to $N^2$ dimensions of the following considerations for positive reals $\zeta_i$, $i = 1, 2, 3$: 
\begin{equation}
	\begin{split}
		(\zeta_1 + \zeta_2 + \zeta_3)^2 &= \zeta_1^2 + \zeta_2^2 + \zeta_3^2 + 2 (\zeta_1 \zeta_2 + \zeta_1 \zeta_3 + \zeta_2 \zeta_3) \\
		&\leq 3(\zeta_1^2 + \zeta_2^2 + \zeta_3^2),
	\end{split}
\end{equation}
since $0 \leq (\zeta_i - \zeta_j)^2 = \zeta_i^2 + \zeta_j^2 - 2 \zeta_i \zeta_j \Leftrightarrow 2 \zeta_i \zeta_j \leq \zeta_j^2 + \zeta_j^2$ for $i,j \in \{ 1, 2, 3\}$.
Therefore, the Lipschitz constant of $\Phi(x,u)$ is given as $\gamma = \sqrt 3 \varrho N + \norm{\tilde A}_{\text F}$, where
\begin{equation}
	\norm{\tilde A}_{\text F} := \sqrt{ \sum_{i = 1}^n \sum_{j = 1}^n (\tilde a_{i,j})^2}. 
	%= \sqrt{2 N^2 \alpha^2} = \sqrt{2} N \alpha.
\end{equation}
The proposed matrix \eqref{eqn:artifical-coupling} is dense and yields a high Lipschitz constant.  
Therefore, only necessary terms in $\tilde A$ are considered such that $(A - \tilde A, C)$ is still observable.
It turns out that the following block structure reduces $\norm{\tilde A}_{\text F}$ and ensures observability, since it introduces the necessary coupling between nitrate and carbon, as well as carbon and microbes:
\begin{equation}
	\tilde A = \begin{bmatrix}
		0_{N^2 \times N^2} & Q^\car & 0_{N^2 \times N^2} & 0_{N^2 \times p} \\
		0_{N^2 \times N^2} & 0_{N^2 \times N^2} & Q^\m & 0_{N^2 \times p} \\
		0_{N^2 \times N^2} & 0_{N^2 \times N^2} & 0_{N^2 \times N^2} & 0_{N^2 \times p} \\
		0_{p \times N^2} & 0_{p \times N^2} & 0_{p \times N^2} & 0_{p \times p}
	\end{bmatrix},
\end{equation}
where $Q^i = \nabla_{C^i} R(C^\nat, C^\car, C^\m)\mid_{x = x_{\text{OP}}}$ and $i \in \{ \car, \m \}$.

\subsection{Approaches to determine the Observer Gain} \label{sec:observer-gain}

Since $\tilde A$ is chosen such that $(A - \tilde A, C)$ is observable, the observer gain $L$ can be computed such that all eigenvalues of the error dynamics  
\begin{equation} \label{eqn:error_dynamics}
	\dot{e} = \left[A - \tilde A - L C\right] e
\end{equation}
are negative, which ensures that the norm of the observation error $\norm{e(t)}$ converges to zero.
The gain is determined via a Riccati inequality.
In this approach, a symmetric, positive definite matrix $P = P^\top \succ 0$ is computed to satisfy
\begin{equation} \label{eqn:Lyap-equ-thau}
	(A - \tilde A - L C)^\top P + P (A - \tilde A - L C) \prec -Q,
\end{equation}
with some positive definite matrix $Q \in \R^{n \times n}$ \eg $Q = I_{n \times n}$.
Obtaining an observer gain $L$ by solving inequality \eqref{eqn:Lyap-equ-thau} relies on the observer of Thau \citep{Thau1973-observing}.
The observation error $\hat x(t)$ converges to zero, if
\begin{equation} \label{eqn:convergence-thau}
	\gamma < \gamma_{\text{max}} := \frac{\lambda_{\text{min}}(Q)}{2 \lambda_{\text{max}}(P)}
\end{equation}
holds, where $\gamma$ is the Lipschitz constant given in \eqref{eqn:Lipschitz_nonlin} and $\lambda_{\text{min}}(A)$, or respectively $\lambda_{\text{max}}(A)$ is the minimal/maximal eigenvalue of the matrix $A$ \citep{Thau1973-observing}.

In contrast to Thau's observer, Raghavan and Hedrick \citep{Raghavan1994-observer} proposed an observer as a solution of the following Riccati inequality
\begin{equation} \label{eqn:Lyap-equ}
	(A - \tilde A - L C)^\top P + P (A - \tilde A - L C) + \gamma^2 P P + I \prec 0.
\end{equation}
%The resulting matrix is used to define a CLF $V(\tilde x) = \tilde x^\top P \tilde x$.
If there exists a matrix $P$ which satisfies \eqref{eqn:Lyap-equ}, then the error dynamics \eqref{eqn:system-highgain-error}
is asymptotically stable \citep{Robenack2012-structure}.
This approach yields the desired observer gain $L$.
In comparison to condition \eqref{eqn:convergence-thau} of Thau's observer, the observer proposed by Raghavan and Hedrick includes a link between the gain $L$ of the linear part and the Lipschitz constant $\gamma$ of the non-linear part of \eqref{eqn:system-ZRD}.
The advantage is, that this link yields a method to design a stable observer instead of only checking asymptotic convergence of the observation error \citep{Raghavan1994-observer}.

%According to \citep{Rajamani1998-observers}, asymptotic stability of the observation error \eqref{eqn:system-highgain-error} is ensured, if $L$ can be chosen to satisfy
%\begin{equation} \label{eqn:rajamani}
%	\gamma < \gamma_{\text{max}} := \sup_L \text{min}_{\omega \geq 0} \sigma_{\text{min}} (A - \tilde A - L C - j \omega I),
%\end{equation}
%where $\sigma_{\text{min}}(K)$ is given as the smallest singular value of the matrix $K$, which is decomposed via singular value decomposition (SVD) \citep{Golub1971-singular}.

\subsection{Model Reduction} \label{sec:reduced-model-disturbance}

For large values of $N$, the LMIs \eqref{eqn:Lyap-equ-thau} and \eqref{eqn:Lyap-equ} become high dimensional.
In case of limited computational capacity, the order of those LMIs has to be reduced such that the solver is still able to handle them.
For this task, we propose to treat the reaction term with concentrations of organic carbon and microbes as disturbance $d(t)$, since their concentrations are not as important to know for fertilization as the concentrations of nitrate on the field.
We assume that the disturbances are generated via the following model:
\begin{equation} \label{eqn:disturbace-generator}
	\begin{aligned}
		\dot x_\text{d} &= A_\text{d} x_\text{d}, \quad x_\text{d}(0) = x_{\text{d},0}, \\
		d &= C_\text{d} x_\text{d}.
	\end{aligned}
\end{equation}

We consider \eqref{eqn:system-ZRD}, and define $A^\nat$ as the $N^2 \times N^2$ block in the upper-left corner of $A$ and $B^\nat$ as the $N^2 \times N$ block in the upper-left corner of $B$.
Therefore, $\dot C^\nat = A^\nat C^\nat + B^\nat C^\nat_\partial$ holds, which describes convection and diffusion of nitrate.
Furthermore, $C^\nat$ is defined as $C$ in \eqref{eqn:system-ZRD}.
The remaining part is the reaction, which is considered as the disturbance $d$. 
To obtain the desired structure of \eqref{eqn:disturbace-generator}, we approximate $\nicefrac{c^i_{j,k}}{(K_i + c^i_{j,k})}$ by $\nicefrac{c^i_{j,k}}{K_i}, \ i \in \{ \nat, \car \}$, which is eligible for small concentrations. 

Furthermore, we assume a 1$^{\text{st}}$-order model for the derivative of $c^i_{j,k}(t), \ i \in \{ \nat, \car, \m \}$, such that its solution is given as $c^i_{j,k}(t) = e^{-\lambda_i t} c^i_{0, j, k}$, $\lambda_i > 0$.
The parameter $\lambda_i$ describes the decrease of a related substance $i \in \{ \nat, \car, \m \}$ and $x^i_{0, j, k}$ is the assumed concentration at point $(j,k)$ at $t = 0$.
Based on different layers of soil, those parameters can be chosen differently.
For the sake of simplicity, we focus on one $\lambda_i$ for nitrate, carbon, and respectively microbes at the whole field.

Including these simplifications in the reaction term, the disturbance reads as
\begin{equation}
	\begin{aligned}
		&d_{(j-1)N + k}(t) = R(c^\nat_{j,k}, c^\car_{j,k},c^\m_{j,k}) \\
		&\approx r_{\text{max}} \frac{c^\nat_{j,k}(t)}{K_\nat} \frac{c^\car_{j,k}(t)}{K_\car} c^\m_{j,k}(t) \\
		&= r_{\text{max}} \frac{e^{-\lambda_\nat t} c^\nat_{0, j, k}}{K_\nat} \frac{e^{-\lambda_\car t} c^\car_{0, j, k}}{K_\car} e^{-\lambda_\m t} c^\m_{0, j, k} \\
		&= \underbrace{\frac{r_{\text{max}}}{K_\nat K_\car} c^\nat_{0, j, k} c^\car_{0, j, k} c^\m_{0, j, k}}_{= c_\text{d}^{(j-1)N + k}} \underbrace{e^{-(\lambda_\nat + \lambda_\car + \lambda_\m) t}}_{= x_\text{d}(t)},
	\end{aligned}
\end{equation}
where $c_\text{d}^{(j-1)N + k}$ is given the $(j-1)N + k$-th element of $C_\text{d}$.
Therewith, the system structure reads as
\begin{equation} \label{eqn:system-disturbance}
	\begin{aligned}
		\begin{pmatrix} \dot{x} \\ \dot{x}_\text{d} \end{pmatrix} &= \begin{pmatrix} A^\nat & C_\text{d} \\ 0 & A_\text{d} \end{pmatrix} \begin{pmatrix} x \\ x_\text{d} \end{pmatrix} + \begin{pmatrix} B^\nat \\ 0 \end{pmatrix} u, \\
		y &= \begin{pmatrix} C^\nat & 0_{p \times 1} \end{pmatrix} \begin{pmatrix} x \\ x_\text{d} \end{pmatrix}
	\end{aligned}
\end{equation}
where 
\begin{equation}
	\begin{aligned}
		C_\text{d} &= \begin{pmatrix}
			r_{\text{max}} K_\nat K_\car \hat c_{0} \\
			0_{p \times 1}
		\end{pmatrix} \in \R^{N^2 + p \times 1}, \\
		\hat c_{0} &= \begin{pmatrix} 
			c^\nat_{0,1,1} c^\car_{0,1,1} c^\m_{0,1,1} \\
			c^\nat_{0,1,2} c^\car_{0,1,2} c^\m_{0,1,2} \\
			\vdots \\ 
			c^\nat_{0,N,N} c^\car_{0,N,N} c^\m_{0,N,N} \end{pmatrix} \\
			A_\text{d} &= - (\lambda^{\nat} + \lambda^\car + \lambda^\m) \in \R.
	\end{aligned}
\end{equation}
The dimension of the matrices are given as $A \in \R^{N^2+p+1 \times N^2+p+1}$, $B \in \R^{N^2+p+1 \times N}$, and $C \in \R^{p \times N^2+p+1}$.

\section{Implementation and case-study} \label{sec:impl-case-study}

\subsection{high-gain observer for the Nitrification Process \eqref{eqn:system-ZRD}} \label{sec:case-study-1}

For the implementation, the grid size $N$ was chosen as $N = 7$, since the computation time as well as the numerical difficulties increase significantly with growing $N$ due to the high-dimensional LMIs.
Furthermore, three sensors were assumed and placed equidistantly to measure the uppermost nitrate concentrations \ie $p = 3$ and $c^\nat_{1,1},c^\nat_{4,1},c^\nat_{7,1}$ are measured.
Therefore, \eqref{eqn:coupling} consists of $n = 3 N^2 + p = 150$ states, $m = 3 N = 21$ inputs and $p = 3$ outputs.
The other parameters are given in Table \ref{tab:values}, where $\frac{\text{mol}}{\text{kgw}}$ means mol per kilogram water \citep{Cussler2009-diffusion, Eltarabily2015-numerical}.
The velocities in y-direction depend on the layer of the soil\ie the velocity of water has a higher magnitude in the upper layer than in the lower layer, which happens due to soil composition.
The minus sign in y-direction results from gravity, whereas the velocity in x-direction is caused if the field is located on a slope.

\begin{table}[h]
	\begin{center}
	\caption{Values for the first simulation Study}
	\label{tab:values}
	\begin{tabular}{|l|l|}
	\hline
		Molecular diffusivity \rule{0pt}{8pt} & $\Gamma = 0.19 \cdot 10^{-4} \, \nicefrac{\mathrm{cm}^2}{\mathrm{s}}$ \\ \hline
		Maximum reaction rate \rule{0pt}{8pt} & $r_{\text{max}} = 1.2 \cdot 10^{-4} \, \nicefrac{1}{\mathrm{s}}$ \\ \hline
		Half velocity constant of nitrate \rule{0pt}{8pt} & $K_\nat = 4.5 \cdot 10^{-3} \, \nicefrac{\mathrm{mol}}{\mathrm{kgw}}$ \\ \hline
		Half velocity constant of carbon \rule{0pt}{8pt} & $K_\car = 6.0 \cdot 10^{-3} \, \nicefrac{\mathrm{mol}}{\mathrm{kgw}}$ \\ \hline
		Velocity of water in $x$ direction \rule{0pt}{8pt} & $v^{\text{x}}_{j,k} = 3.5 \cdot 10^{-3} \, \nicefrac{\mathrm{cm}}{\mathrm{s}}$	\\\hline
		Velocity of water in $y$ direction \rule{0pt}{8pt} & $v^{\mathrm{y}}_{j,k}$ \\
		 & $= \begin{cases} -0.2 \, \nicefrac{\mathrm{cm}}{\mathrm{s}}, & \mathrm{ if} \ k < N/2 \\ -0.1 \, \nicefrac{\mathrm{cm}}{\mathrm{s}} &\mathrm{otherwise} \end{cases}$ \\ \hline
		Grid points in one direction \rule{0pt}{8pt} & $N = 7$ \\ \hline	
		Distance between points in grid \rule{0pt}{8pt} & $\Delta = 10 \, \mathrm{cm}$ \\ \hline
		Time constant of nitrate sensor \rule{0pt}{8pt} & $\tau_\nat = 1 \, \mathrm{s}$ \\ \hline
		Initial value of system & $x_0^\top = \begin{pmatrix} 10^{-4} &\cdots & 10^{-4}\end{pmatrix}$ \\ \hline
		Initial value of observer & $\hat x_0^\top = 3 \cdot \begin{pmatrix} 10^{-4} &\cdots & 10^{-4}\end{pmatrix}$ \\ \hline
	\end{tabular}
	\end{center}
\end{table}

The observer gain $L$ was computed in MATLAB via an LMI solver, where the quadratic terms in \eqref{eqn:Lyap-equ-thau} and \eqref{eqn:Lyap-equ} are removed via Schur's complement \citep{Robenack2012-structure}.
Based on the computed $L$, the upper bound for the Lipschitz constant was determined via \eqref{eqn:convergence-thau} as $\gamma_{\mathrm{max}} = 1.7 \cdot 10^{-4}$, whereas \eqref{eqn:Lipschitz_nonlin} yields a Lipschitz constant $\gamma = 1.2 \cdot 10^{-5}$, which is less than $\gamma_{\mathrm{max}}$.
Thus, the error dynamics \eqref{eqn:system-highgain-error} converge asymptotically to zero.
The Lipschitz constant $\varrho$ of the reaction term in \eqref{eqn:Lipschitz_nonlin_proof2} was identified numerically as $\varrho = 1.4 \cdot 10^{-8}$ for the given values.
The operating point for the linearization was chosen as $c_{j,k,\mathrm{OP}}^\nat = 5 \cdot 10^{-3}$, $c_{j,k,\mathrm{OP}}^\car = 3 \cdot 10^{-3}$, and $c_{j,k,\mathrm{OP}}^\m = 2 \cdot 10^{-4}$ to ensure observability of the linear system (cf. Section \ref{sec:observ}).
Also disturbances can be included in the system description with a magnitude $\gamma_{\mathrm{dist}}$. 
If $\gamma + \gamma_{\mathrm{dist}} < \gamma_{\text{max}}$ still holds, then exponential convergence of the observation error (OE) is not endangered.

\begin{figure}[!h]
	\centering
  	\includegraphics[width = 0.45\textwidth]{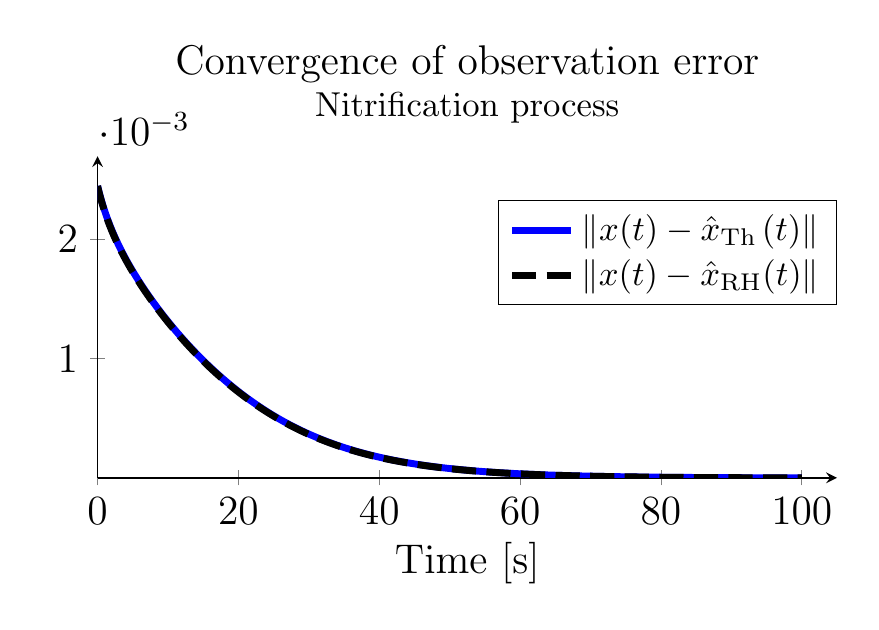}
	\caption{Convergence of the observation error for high-gain observers of Thau and Raghavan/Hedrick based on the nitrification process as a model.}
	\label{fig:results}
\end{figure}

The convergence of the observation error $\norm{x(t) - \hat x_{\cdot}(t)}$ is shown in Fig. \ref{fig:results} for both high-gain observers.
The estimated states based on Thau's observer are denoted as $\hat x_{\mathrm{Th}}(t)$, whereas $\hat x_{\mathrm{RH}}(t)$ denotes the estimated states based on the observer of Raghavan and Hedrick.
In this case, they yield nearly the same result, whereas the OE of Thau's observer is about $10^{-5}$ higher than the other one.
Both OEs decline exponentially.
As an input, a step with height $10^{-2}$ on the nitrate concentrations was given on the system to simulate fertilization\ie $u = 10^{-2} \cdot \begin{pmatrix} 1_{1 \times 7} & 0_{1 \times 14} \end{pmatrix}^\top$, where $1_{1 \times n} = \begin{pmatrix} 1 & \cdots & 1 \end{pmatrix}^\top \in \R^n$.

The number of grid points can be increased for higher computational powers.
Since the size of the system grows quadratically with $N$, the resulting LMI becomes very large.
Further investigations are necessary at this point to handle this system especially in view of numerical analysis.
Nevertheless, since all concentrations of the sparse grid are known, the concentrations in between can be interpolated via known methods, such as spline interpolation \citep{deBoor1978-practical}.
In the next section, the reduced order model is considered to increase the number of grid points.

\subsection{high-gain observer for the Reduced Model \eqref{eqn:system-disturbance}} \label{sec:case-study-2}

In this case, the reaction term is considered as a disturbance (cf. Section \ref{sec:reduced-model-disturbance}).
The parameters are the same as in Section \ref{sec:case-study-1}, except for $N$, which is chosen as $N = 11$ for the following computations such that the matrices of the reduced model given as $A \in \R^{125 \times 125}, B \in \R^{125 \times 11}, C \in \R^{3 \times 125}$ have a similar size as in Section \ref{sec:case-study-1}.
The additional parameters are listed in Table \ref{tab:values2}, whereas the Lipschitz constant can be interpreted as a bound for a disturbance.

\begin{table}[h]
	\begin{center}
	\caption{Additional or changed values for the second simulation Study}
	\label{tab:values2}
	\begin{tabular}{|l|l|}
	\hline
		Grid points in one direction \rule{0pt}{8pt} & $N = 11$ \\ \hline	
		Decrease of disturbance & $\lambda_\nat + \lambda_\car + \lambda_\m = 6.1 \cdot 10^{-3}$ \\ \hline
		Initial value of disturbances & $\hat c_{j,k} = 0.3 \cdot 10^{-9}$ \\ \hline
		Lipschitz constant & $\gamma = 1 \cdot 10^{-3}$ \\ \hline
	\end{tabular}
	\end{center}
\end{table}

\begin{figure}[!h]
	\centering
  	\includegraphics[width = 0.45\textwidth]{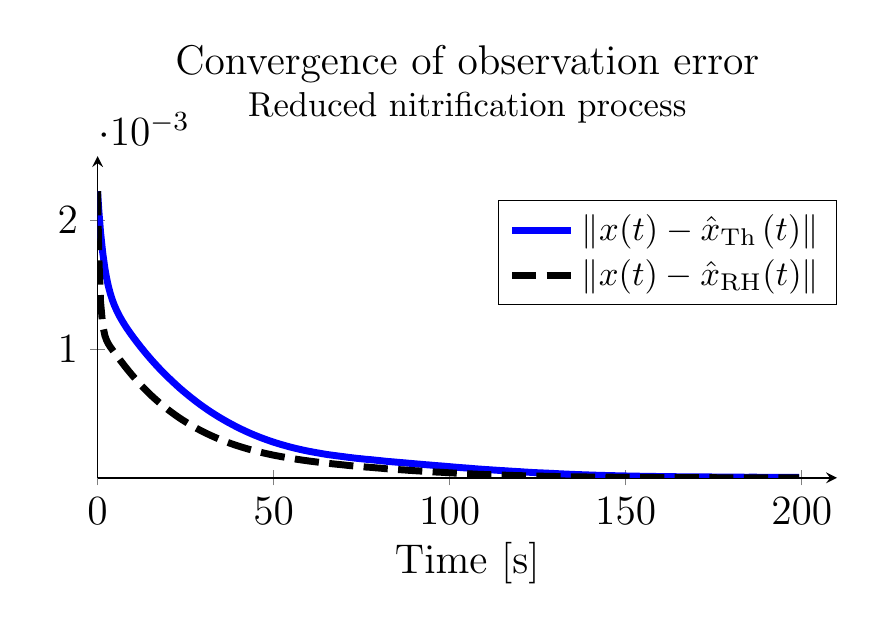}
	\caption{Convergence of the observation error for high-gain observers of Thau and Raghavan/Hedrick based on the reduced nitrification process as a model.}
	\label{fig:results2}
\end{figure}

Figure \ref{fig:results2} shows that the observation error (OE) converges exponentially to zero for both high-gain observers (HGO).
Note that $x(t)$ includes only the states of the nitrate concentrations, since $x_\mathrm{d}(t)$ slows down the convergence because these additional states converge slowly to zero.
Nevertheless, their value is not crucial to know.
If Thau's HGO is considered, the OE is smaller than $10^{-4}$ after approx. $95$ seconds, whereas the OE of Raghavan and Hedrick's improved observer enters this vicinity already after $70$ seconds.
Therefore, it significantly accelerates the convergence of the OE.

%\section{\red{Conclusion}} \label{sec:conclusion}
\section{Conclusion and outlook} \label{sec:concl-outlook}

The novelty of this paper is the application of the two high-gain observers to the nitrification process including sensor measurement dynamics.
The considered nitrification process was modeled based on the advection-diffusion-reaction equation.
The dynamics provided by the sensors were included to the model, which generates a set of partial differential equations.
After discretizing the system on a grid, the resulting model is described as a set of ordinary differential equations.
This set of equations formed the basis for further considerations, namely the high-gain observer.
The system was decomposed in a linear and a nonlinear part.
It turned out that the linear part is not observable due to the block-diagonal structure of the system matrix $A$.
An additional coupling based on the linearization was introduced to ensure observability.
Afterwards, the nonlinearity was bounded by a Lipschitz constant.
The observer gain was computed via two slightly different approaches based on the Riccati inequality, which results in high-gain observers proposed by Thau \eqref{eqn:Lyap-equ-thau} and by Raghavan/Hedrick \eqref{eqn:Lyap-equ}.
They ensure an asymptotically converging observation error, but are characterized by a high computational burden, especially for large systems.
Therefore, a model reduction was proposed, where the reaction term is treated as a disturbance, which allows a higher number of simulation points.
Asymptotic stability of the estimation error was proven theoretically and in a case-study for both proposed approaches.
Thus it was shown, that the high-gain observer estimates the desired concentrations based on the measurements.
The results can be substantiated via experiments on a field with real data.
This remains a task for the future.
However, there are some properties that have to be considered in the model.
Those might be time-varying velocities of water, uncertainties caused by wear of sensor materials, or different sensor measurement dynamics, such as second-order model, or electrochemical properties of the sensor.
Furthermore, the number of simulation points can be increased, which requires a closer look on the algorithm to avoid numerical problems.
If these computational burdens are eliminated, one might obtain results for a higher number of grid points or even values for a three dimensional grid.
The contribution of publication focuses on the method rather than a high-dimensional calculation example.
Also practical issues like model uncertainty, i.e. parameter uncertainty, model structural mismatch, or soil heterogeneity can be considered in further works to provide a more realistic setup.
In this case, a closer look on sensitivity of the high-gain observer to measurement noise has to be made.
However, existing approaches provide a robustness to measurement errors \citep{Ahrens2009-high}.

All these preliminary investigations can be used in a further control setup, where strategies for the improved usage of fertilizer can be derived based on the estimated concentrations.
Therefore, the approach serves as an important step to determine the current concentrations of substances on a field based on a few measurements.
This would avoid all the discussed negative effects of overfertilization.

%\begin{ack}
%This work was partially funded by the European Union, European Social Found ESF, Saxony.
%\end{ack}

\bibliography{bib}             % bib file to produce the bibliography
                                                     % with bibtex (preferred)

\end{document}